\documentclass[
    ,final            
  ]
  {aipproc}

\layoutstyle{8x11single}
\graphicspath{{./}}

\begin{document}

\title{First Results with Heavy-Ion Collisions at LHC from ALICE}


\author{Domenico Elia$^a$ for the ALICE Collaboration}{
  address={$^a$INFN Sezione di Bari, via Orabona 4, 70126 Bari, Italy}
}

\begin{abstract}
In November 2010 the ALICE experiment at CERN has collected the first Pb--Pb
collisions at $\sqrt{s_{NN}}$ = 2.76 TeV produced by the LHC. 
A first characterization of the hot and dense state of matter produced in this 
new energy domain
became available shortly after the run. In this paper we present the results
on charged-particle multiplicity, Bose-Einstein correlations, elliptic flow and 
their dependence on the collision centrality. Results from first measurements of 
strange and identified particle production and suppression of high-momentum hadrons
with respect to $pp$ collisions are also reported. 
\end{abstract}

\maketitle


\section{INTRODUCTION}

ALICE (A Large Ion Collider Experiment) has been specifically designed to
study the properties of the strongly interacting matter created in
heavy-ion collisions at the LHC energies\cite{ppr1,ppr2}.
The experimental apparatus is optimized to measure a large variety of observables in the
very high multiplicity environment of the heavy-ion collisions.
The detector is also recording $pp$ collisions as required primarily to have comparison data 
for the heavy-ion programme \cite{ppr1}.
It consists of a central part ($|\eta|$~$<$~0.9) to detect hadrons, 
electrons and photons, a forward spectrometer to measure muons and additional smaller forward 
detectors for event characterization and triggering: a detailed description of the eighteen subsystems
can be found in \cite{technote}. The detector is fully installed, commissioned and operational, with 
the exception of the Transition Radiation Detector (TRD) and the ElectroMagnetic CALorimeter (EMCAL):
both systems had a 40\% of their active area installed in 2010, the EMCAL has been fully
installed in the winter shutdown 2010/11 and the TRD will be completed for the 2012 data taking. 
\par
After collecting data from $pp$ collisions at $\sqrt{s_{NN}}$~=~0.9, 2.36 and 7 TeV, the first Pb--Pb
collisions were recorded in November 2010 at a center-of-mass energy per nucleon pair of 2.76 TeV,
more than a factor 10 higher compared to the previous heavy-ion experiments. 
In the following, after a brief description of the data taking conditions, the very first
results with Pb--Pb collisions published by ALICE will be presented and discussed.

\section{DATA ANALYSIS AND RESULTS}

During the first heavy-ion data taking period at LHC, from 4 to 114 bunches of about
7~$\times$~10$^7$ ions of $^{208}$Pb were brought into collision at $\sqrt{s_{NN}}$~=~2.76 TeV
in the ALICE interaction region. The maximum rate of hadronic events was about 100~Hz,
corresponding to an estimated luminosity of about 2~$\times$~10$^{25}$~cm$^{-2}$~s$^{-1}$
reached at the end of the run. Signals from the outermost pixel layer (SPD, Silicon Pixel
Detector, $|\eta|$~$<$~1.5) and from two forward scintillator hodoscopes 
(VZERO-A and VZERO-C, 2.8~$<$~$|\eta|$~$<$~5.1 and -3.7~$<$~$|\eta|$~$<$~-1.7)
were used to implement a minimum-bias interaction trigger: a total of about 30 M minimum-bias
Pb--Pb interactions to tape were finally recorded.
The sum of the amplitudes of the signals in the VZERO counters is also used as a measure
of the event centrality.
\par
\begin{figure}[htb]
\hspace{1.7cm}
\includegraphics[scale=0.375]{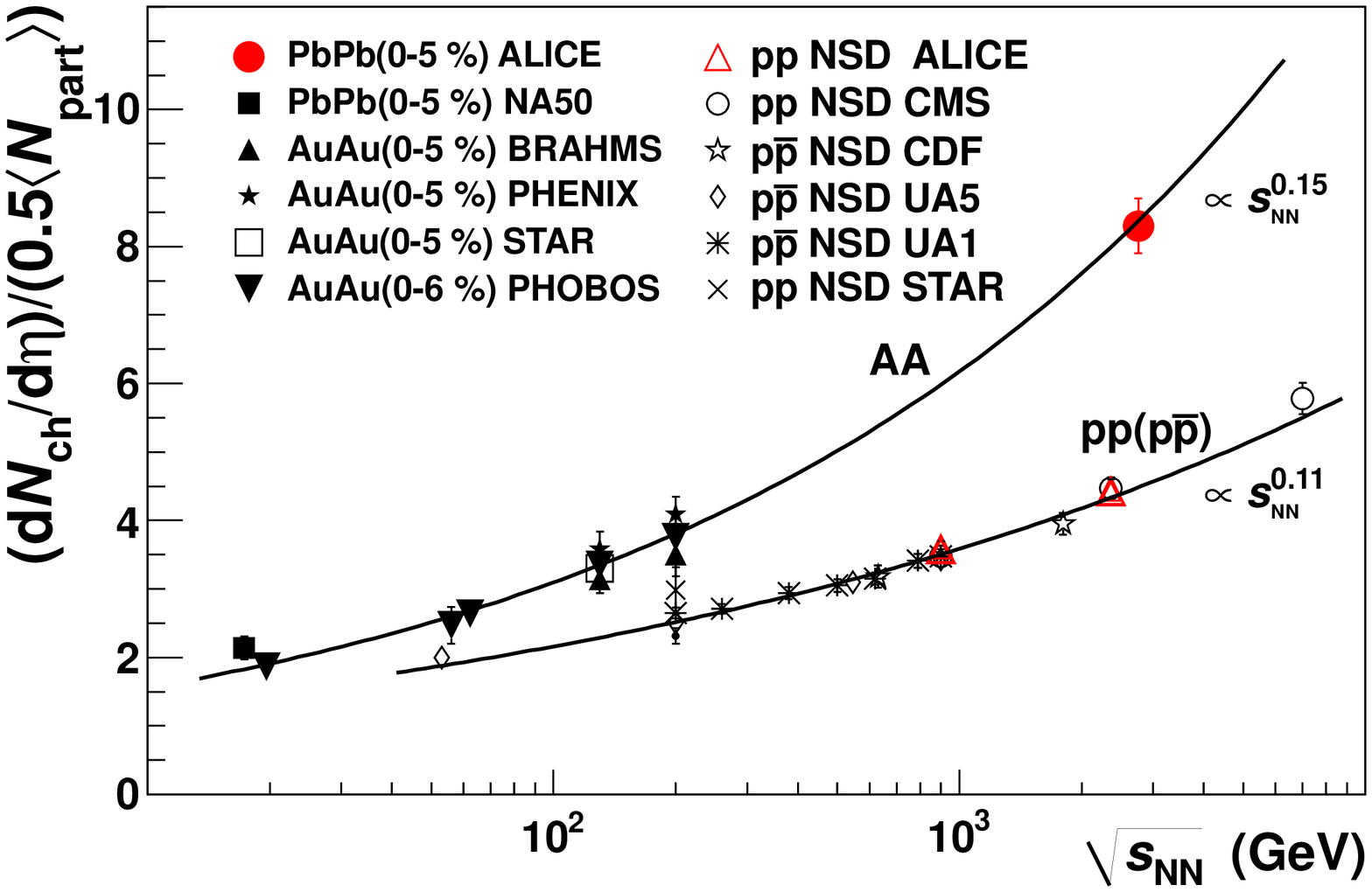}
\hspace{\fill}
\begin{minipage}[t]{100mm}
\hspace{0.8cm}
\includegraphics[scale=0.375]{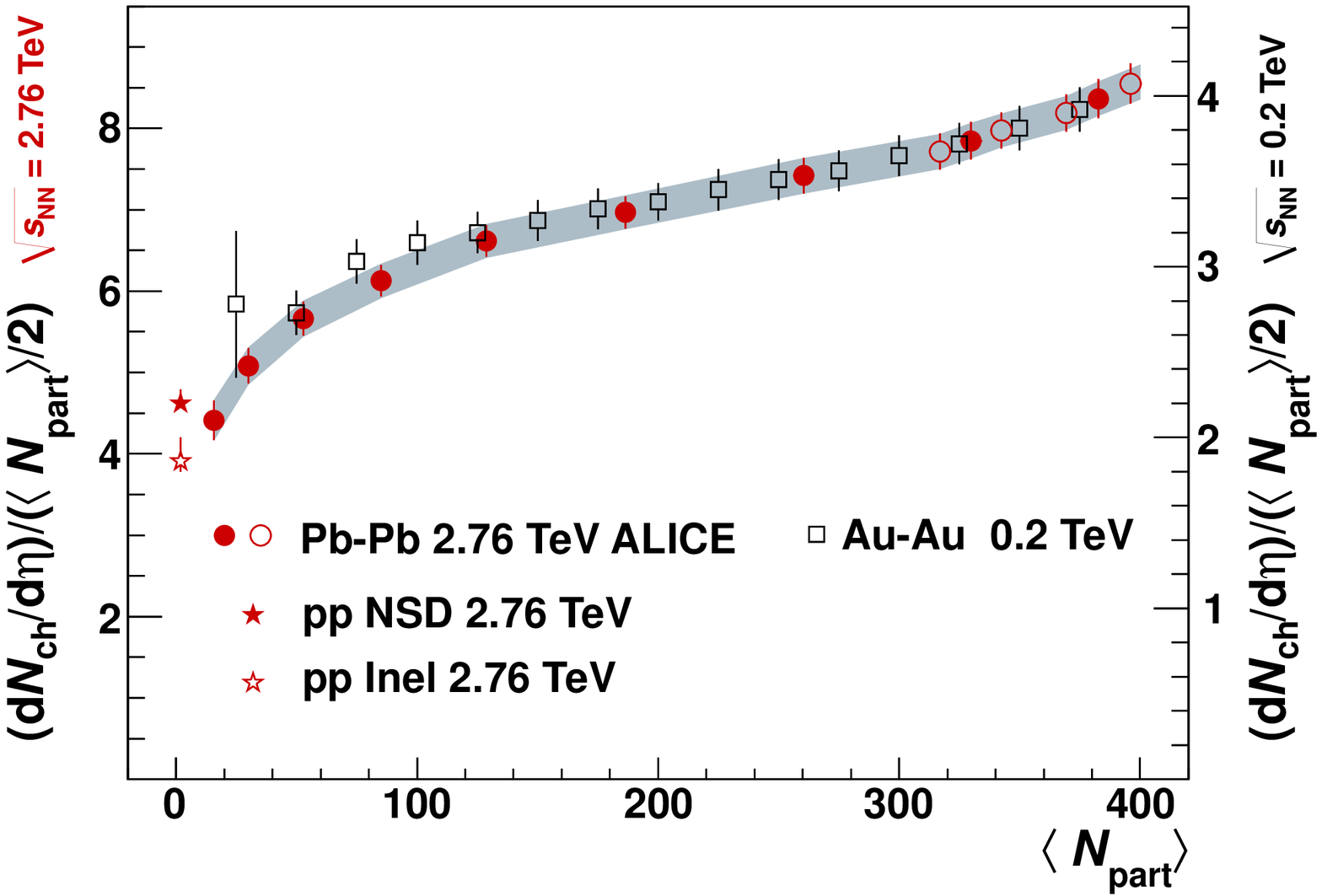}
\end{minipage} 
\caption{Left panel: Charged-particle pseudorapidity density per participant pair
for central nucleus-nucleus and non-single diffractive $pp$ ($p\overline{p}$)
collisions as a function of $\sqrt{s_{NN}}$.  
Righ panel: Centrality dependence of charged-particle pseudorapidity density per participant pair
for Pb--Pb collisions at $\sqrt{s_{NN}}$ = 2.76 TeV and
Au--Au collisions at $\sqrt{s_{NN}}$ = 0.2 TeV. The scale for for lower-energy data is shown
on the righ-hand side and differs from that at higher energy by a factor 2.1.}
\label{mult}
\end{figure}

The first measurement concerned the charged-particle multiplicity density in
central collisions. It was based on the counting of the SPD tracklets, i.e. combinations of pairs 
of hits on the two pixel layers aligned to the main interaction vertex within predefined
tolerances. The value $dN_{ch}/d\eta$~=~1584~$\pm$~4(stat)~$\pm$~76(syst) 
measured in the 5\% most central Pb--Pb collisions at $\sqrt{s_{NN}}$~=~2.76 TeV corresponds to 
a value of 8.3~$\pm$~0.4(syst), with negligible
statistical error, when normalized to the number of participant pairs. 
In the left panel of Fig.\ref{mult} the ALICE
result in Pb--Pb collisions is compared with other measurements in nucleus-nucleus
and non-single diffractive $pp$ ($p\overline{p}$) collisions in a wide range of energies \cite{papmult1}.
The pseudorapidity density increases significantly, by a factor 2.2, when going from 
RHIC highest energy Au--Au collisions to the Pb--Pb collisions at 2.76 TeV.
An estimate of the Pb--Pb energy density $\epsilon$ at LHC indicates approximately
a factor 2.5 increase with respect to that at RHIC, assuming identical equilibration time at both energies. 
The average multiplicity per participant pair is also found to be a factor 1.9 higher 
than in $pp$ and $p\overline{p}$ collisions at similar energies.  
The measurement of $dN_{ch}/d\eta$ per participant pair as a function of centrality in 
Pb--Pb collisions at $\sqrt{s_{NN}}$~=~2.76 TeV, illustrated in the right panel of
Fig.\ref{mult}, shows a steady increase by about a factor 2 between peripheral and 
central collisions, with a behaviour similar to that observed in Au--Au collisions 
at $\sqrt{s_{NN}}$~=~0.2~TeV \cite{papmult2}.

\begin{figure}[htb]
\hspace{1.7cm}
\includegraphics[scale=0.95]{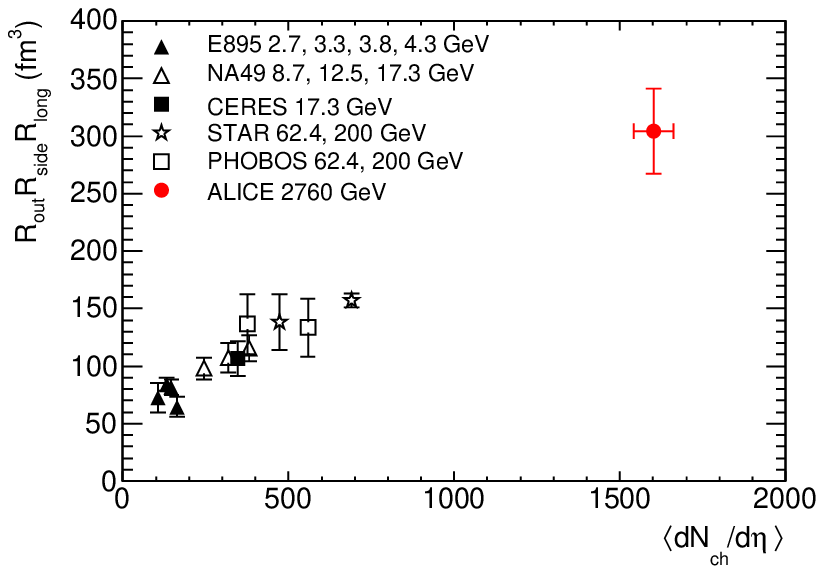}
\hspace{\fill}
\begin{minipage}[t]{100mm}
\hspace{0.2cm}
\includegraphics[scale=0.95]{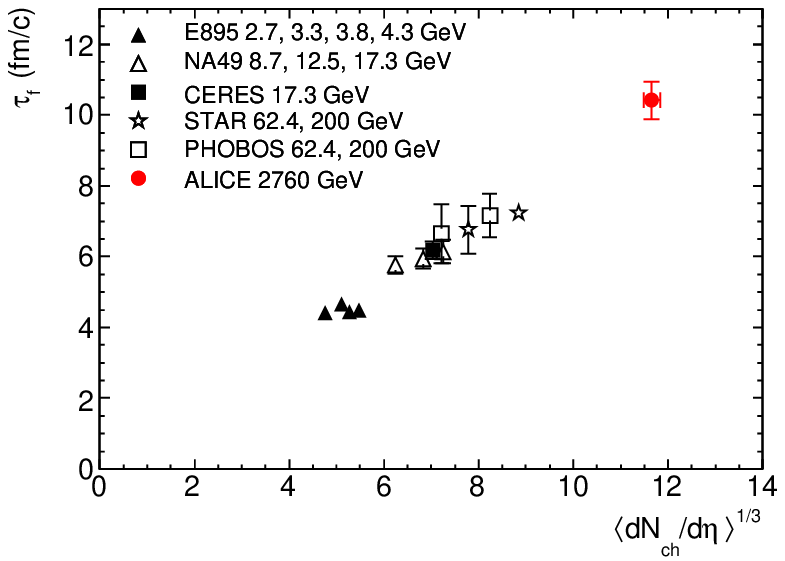}
\end{minipage} 
\caption{Product of the three pion HBT radii (left panel) and decoupling time (right panel):
the ALICE points (red filled dots) are compared to the results obtained by other
experiments at lower energies.}
\label{hbt}
\end{figure}

Results concerning the freeze-out volume and the total lifetime of the 
system created in Pb--Pb collisions at $\sqrt{s_{NN}}$~=~2.76 TeV have been obtained
by measurement of identical particle interferometry (HBT) \cite{paphbt}. 
The product of the three radii, presented in the left panel of 
Fig.\ref{hbt}, is connected to the volume of the homogeneity region
by a factor of (2$\pi$)$^{3/2}$.
The volume exibits a linear dependence on the charged-particle pseudorapidity density 
and reaches a value of about 4600 fm$^3$ for central Pb--Pb collisions, nearly 5 times
the volume of a Pb nucleus. 
In the righ panel of Fig.\ref{hbt} the same trend can be observed for the extracted 
decoupling times: here the ALICE estimate is about 30\% higher compared to RHIC.

\begin{figure}[hbt]
\hspace{2.5cm}
\includegraphics[scale=0.38]{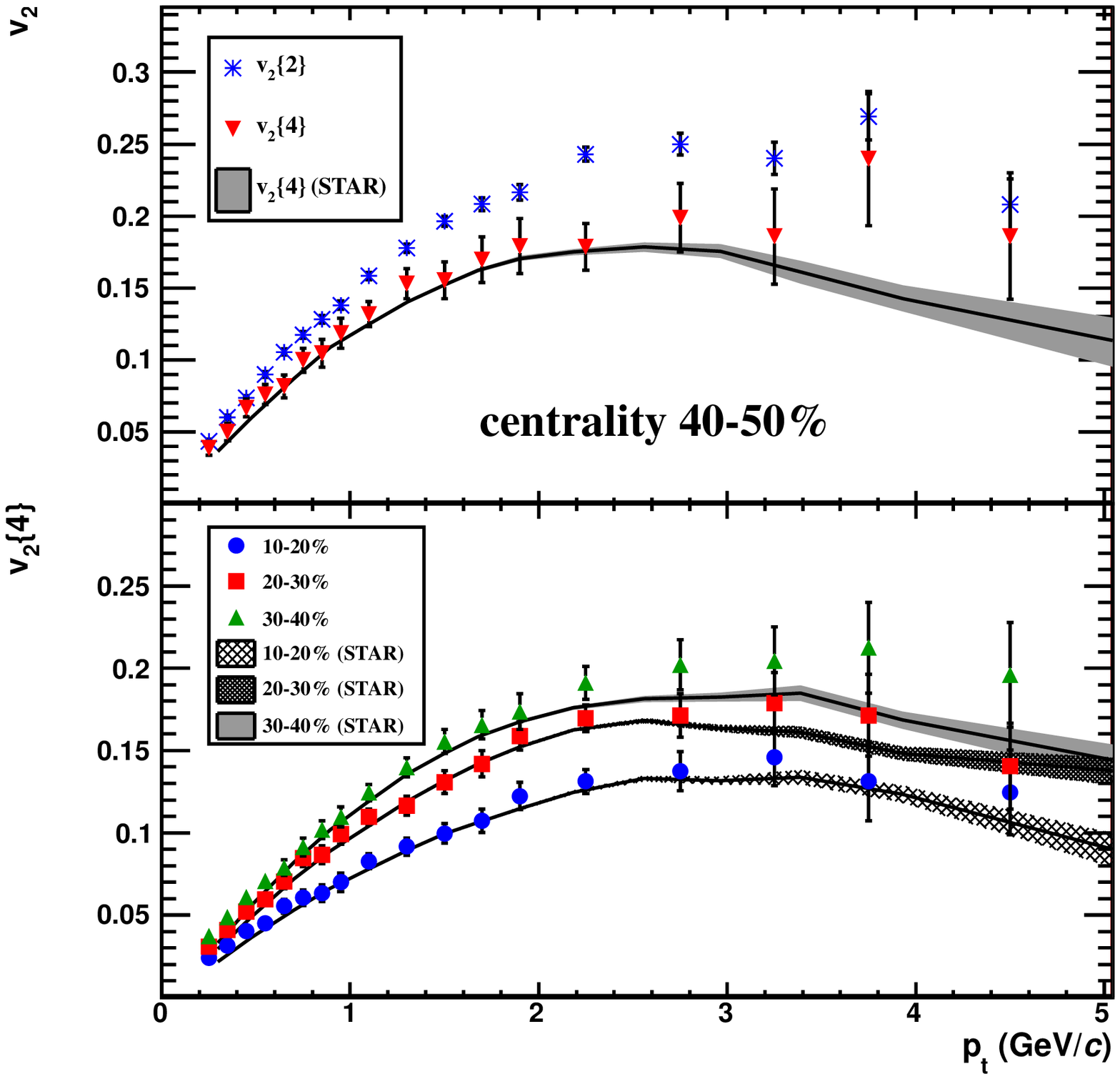}
\hspace{\fill}
\begin{minipage}[t]{100mm}
\hspace{0.5cm}
\includegraphics[scale=0.297]{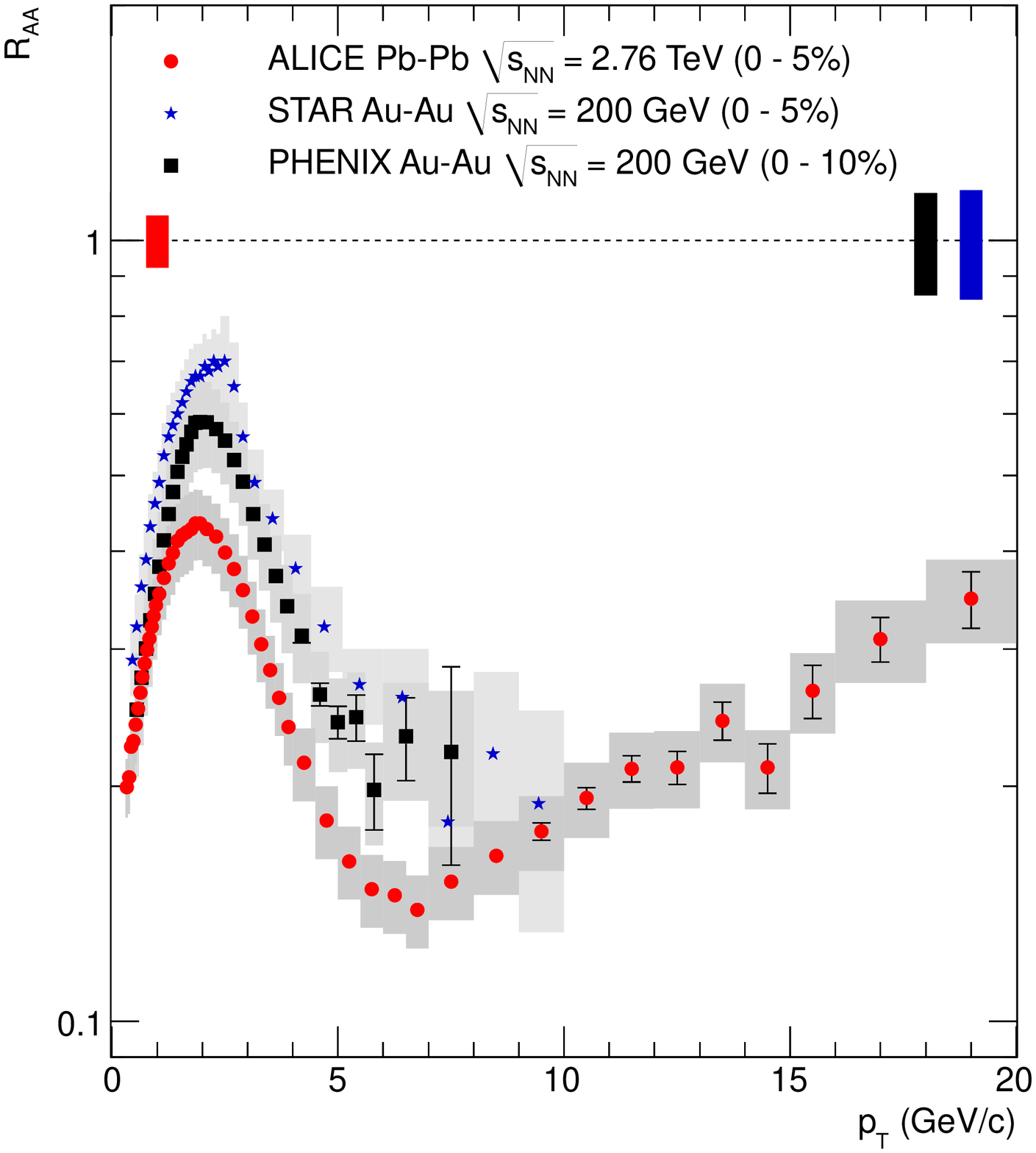}
\end{minipage} 
\caption{Left panel: v$_2$(p$_T$) for centrality 40-50\%, from 2- and 4-particle
cumulant methods for Pb--Pb collisions at LHC and Au--Au collisions at $\sqrt{s_{NN}}$~=~0.2 TeV (top)
and v$_2$\{4\}(p$_T$) for various centrality classes compared with STAR measurements (bottom).
Right panel: $R_{AA}$ in central Pb--Pb collisions at $\sqrt{s_{NN}}$~=~2.76 TeV
compared to measurements at RHIC.}
\label{flowhighpt}
\end{figure}

Elliptic flow (v$_2$) measurements indicate the creation of a strongly-interacting
medium of low viscosity in nucleus-nucleus collisions at RHIC energies. 
The v$_2$ integrated over p$_T$ as measured by ALICE in
Pb--Pb collisions at $\sqrt{s_{NN}}$~=~2.76 TeV has been compared with results at
lower energies \cite{papflow}. An increase of about 30\% with respect to the top RHIC 
energy is observed: this confirms that the medium created in Pb--Pb collisions at LHC
behaves very much as at RHIC and should constrain the temperature
dependence of the ratio $\eta$/s between shear viscosity and entropy density. 
The left panel of Fig.\ref{flowhighpt} shows the p$_T$-differential flow for various 
centralities, obtained with different methods and compared with
the corresponding results in Au--Au collisions at $\sqrt{s_{NN}}$~=~0.2 TeV:
the value of v$_2$(p$_T$) does not change within the uncertainties from RHIC to LHC
and the transverse momentum dependence is qualitatively similar for all centralities.
In particular, the result that the differential elliptic flow stays the same
over almost two order of magnitudes in energy was not anticipated:
the increase in the integrated v$_2$(p$_T$) between RHIC and LHC is due to
an increase in the average transverse momentum, which can be partly attributed to 
an increased radial flow \cite{papstrangeid}. 
\par
The nuclear modification factor $R_{AA}$(p$_T$) for central Pb--Pb collisions
has been also measured by ALICE with the first heavy-ion data \cite{paphighpt}. 
The measured charged-particle spectra in $|\eta|$~$<$~0.8 and 0.3~$<$~p$_T$~$<$~20~GeV/$c$,
when compared to the expectation in $pp$ at the same $\sqrt{s_{NN}}$ scaled by
the number of underlying nucleon-nucleon collisions, shows a pronounced minimum
around 6-7 GeV/$c$ and then increases significantly at larger p$_T$. 
As shown in the right panel of Fig.\ref{flowhighpt}, the high-p$_T$ suppression measured by ALICE 
is significantly larger than that observed at lower energy, indicating a stronger parton
energy loss in the very dense medium formed in central Pb--Pb collisions at LHC.

Measurements of transverse momentum spectra of identified pions, kaons and protons 
for both charged states have been carried out as well: a stronger power law dependence 
than observed at RHIC suggests a stronger radial flow at LHC.
The ``baryon anomaly" (enhanced baryon to meson ratios) observed at RHIC has been found
also at LHC. The $\Lambda/K^0_s$ ratio measured by ALICE is slightly larger than
that at RHIC, with a very little shift of its maximum in p$_T$ \cite{papstrangeid}.

\section{CONCLUSIONS}

The first ALICE measurements with Pb--Pb collisions at $\sqrt{s_{NN}}$~=~2.76 TeV have been presented. 
In general a smooth evolution from RHIC to LHC has been observed, with qualitatively similar
but quantitatively different results on most of the measurements. While a lot of physics analyses 
are currently ongoing on this first data, ALICE looks forward to the next (higher luminosity)
heavy-ion run to continue the exciting travel into the ``hot and dense matter" Wonderland.
\\

\noindent {\bf Acknowledgements}  Copyright CERN for the benefit of the ALICE Collaboration.

\end{document}